\documentclass[%
preprint,
nofootinbib,
 amsmath,amssymb,
 aps,
 prl,
showkeys
]{revtex4-1}

\usepackage{graphicx}
\usepackage{dcolumn}
\usepackage{bm}
\usepackage{hyperref}

\usepackage{comment}

\newcommand\fig[1] {Fig.\,{\ref{#1}}}

\newcommand\Tab[1] {Tab.\,{\ref{#1}}}

\def\beq{\begin{equation}}
\def\eeq{\end{equation}}
\def\bsp#1\esp{\begin{split}#1\end{split}}
\def\bal#1\eal{\begin{align}#1\end{align}}
\newcommand{\as}{\ensuremath{\alpha_\textrm{s}}}
\newcommand{\cusp}{\text{cusp}}
\newcommand{\mMDT}{\text{mMDT}}
\newcommand{\rL}{\text{L}}
\newcommand{\rR}{\text{R}}
\newcommand{\rd}{\text{d}}
\newcommand{\rg}{\text{g}}
\newcommand{\zcut}{\ensuremath{z_\text{cut}}}

\newcommand\NNLO{N$^2$LO}
\newcommand\NnLL[1]{N$^{#1}$LL}
\newcommand\MCCSM{{\tt MCCSM}}

\begin{document}


\title{Groomed jet mass at high precision}

\author{Adam Kardos}
 \email{kardos.adam@science.unideb.hu}
 \affiliation{University of Debrecen, 4010 Debrecen, PO Box 105, Hungary}
\author{Andrew J.~Larkoski}%
 \email{larkoski@reed.edu}
\affiliation{Physics Department, Reed College, Portland, OR 97202, USA}%

\author{Zolt\'an Tr\'ocs\'anyi}
 \email{Zoltan.Trocsanyi@cern.ch}
 \homepage{http://pppheno.elte.hu}
\affiliation{
 Institute for Theoretical Physics, ELTE E\"otv\"os Lor\'and University, 
Pázmány Péter 1/A, H-1117 Budapest, Hungary
\\ and
MTA-DE Particle Physics Research Group, University of Debrecen, 
4010 Debrecen, PO Box 105, Hungary
}%

\date{\today}

\begin{abstract}
\noindent We present predictions of the distribution of groomed heavy jet mass in electron-positron collisions at the next-to-next-to-leading order accuracy matched with the resummation of large logarithms to next-to-next-to-next-to-leading logarithmic accuracy.  Resummation at this accuracy is possible through extraction of necessary two-loop constants and three-loop anomalous dimensions from fixed-order codes.
\end{abstract}


\maketitle



High-energy electron-positron collisions are considered as ideal tools 
for precision studies of particle interactions. The initial state of the 
hard scattering event is colorless and known precisely, which eliminates 
significant sources of uncertainties that are ubiquitous at hadron colliders 
such as the LHC. For instance, the study of hadronic final states at the 
Large Electron-Positron collider (LEP) was used extensively to study the dynamics 
of strong interactions \cite{Decamp:1990cg,Adeva:1990nu,Adeva:1991nv,Abreu:1992yc,Adriani:1992gs,Decamp:1992wz,Acton:1993zh,Abreu:1999rc,Abbiendi:2001qn,Heister:2002tq,Abbiendi:2004qz} and especially to determine the strong coupling \as.
Yet, the current state of the art does not support these expectations. 
Hence, it is somewhat disappointing that presently the second largest 
spread and uncertainty of determination of \as\ among seven sub-fields is found 
in the group of results based on jets and event shapes of hadronic final states 
in electron-positron annihilation \cite{Tanabashi:2018oca}. This failure of fulfilling 
expectations calls for an investigation of the possible sources.

The comparison of event shape distributions obtained from data collected 
by the LEP experiments and from theoretical predictions obtained in QCD 
perturbation theory reveal the possible causes of such a failure \cite{Salam:2017qdl,Trocsanyi:2019irv}: 
(i) the QCD radiative corrections are large,
(ii) the hadronization corrections are not well understood from first principles,
(iii) the two-types of corrections are strongly anti-correlated for analytic models of hadronization. 
As a result the systematic theoretical uncertainties are large. 
In order to decrease these corrections, one has to select the observables used 
for \as-extraction carefully. For instance, jet rates are expected to be less 
sensitive to hadronization corrections than event shapes \cite{Dokshitzer:1995qm}, which is 
supported by a recent Monte Carlo evaluation, resulting in a competitive 
value for \as\ \cite{Verbytskyi:2019zhh}. The latter study is based on the highest perturbative
order available for two-jet rates: next-to-next-to-next-to-leading order (N$^3$LO) matched 
with the resummation of the first three largest logarithms at all orders (\NnLL{2}) 
in perturbation theory. 

For precision extraction of the strong coupling the logarithmic accuracy should 
extend to next-to-next-to-next-to-leading 
logarithmic order (\NnLL{3}) 
that allows for simple additive matching to fixed-order at \NNLO.  
Such matched predictions are available for thrust \cite{Becher:2008cf}
and $C$-parameter \cite{Hoang:2014wka}, and were used for the extraction 
of \as\ from LEP data \cite{Abbate:2010xh,Hoang:2015hka}. 
However, even so high perturbative accuracy does not guarantee small uncertainty
for the determination of \as\ due to lack of good control over the hadronization.
One way out is to reduce the latter effect. The analysis techniques broadly 
referred to as jet grooming have been introduced to mitigate contamination radiation
in jets from outside of the jet. Jet groomers identify such emissions 
in the jet and remove them from consideration.  The modified mass-drop tagger (mMDT) \cite{Dasgupta:2013ihk,Dasgupta:2013via} 
and soft drop \cite{Larkoski:2014wba} algorithms are the best understood groomers,
due to their unique feature of elimination of non-global logarithms (NGLs)
\cite{Dasgupta:2001sh} that are the leading correlations between in-jet 
and out-of-jet scales. Soft drop was indeed found to reduce the hadronization 
corrections for event shapes in electron-positron annihilation \cite{Baron:2018nfz}.

In this Letter, we present theoretical predictions for the mMDT groomed jet mass in $e^+e^-$ collisions at \NNLO\ matched with \NnLL{3} accuracy in perturbation theory.  Resummation at this accuracy is made possible by the factorization theorem for jet grooming from Ref.~\cite{Frye:2016aiz} and recent extraction of necessary constants and anomalous dimensions at two- and three-loop order \cite{Bell:2018vaa,Bell:2018oqa,Bell:2020yzz,Kardos:2020ppl}.  A demonstration of reduction of scale uncertainties and good convergence of the perturbation series will be presented here, but we leave a detailed study of scale variations and inclusion of non-perturbative corrections to groomed jets established in Ref.~\cite{Hoang:2019ceu} for future work.


The modified mass-drop tagger groomer (mMDT) \cite{Dasgupta:2013ihk}, or soft 
drop with angular exponent $\beta = 0$ \cite{Larkoski:2014wba}, proceeds as follows:
\begin{enumerate}
\itemsep=-2pt
\item Divide the final state of an $e^+e^-\to$ hadrons event into two 
hemispheres in any infrared and collinear safe way.
\item Define a clustering metric $d_{ij}$ between particles $i$ and $j$ 
in the same hemisphere. The metric appropriate for $e^+e^-$ collisions is
\begin{equation}
d_{ij} = 1-\cos\theta_{ij}\,,
\end{equation}
with $\theta_{ij}$ being the angle between the trajectory of the particles.
\item In each hemisphere, apply the Cambridge/Aachen jet algorithm
\cite{Dokshitzer:1997in,Wobisch:1998wt} to produce an angular-ordered 
pairwise clustering history of particles.
\item Starting with one of the hemispheres (say left) and at widest angle, 
step through the Cambridge/Aachen particle branching tree.
At each branching in the tree, test if
\begin{equation}
\frac{\min[E_i,E_j]}{E_i+E_j} > \zcut
\label{eq:groomingtest}
\end{equation}
is satisfied, where $i$ and $j$ are the daughter particles at that branching 
and $\zcut$ is some fixed numerical value where $0\leq \zcut < 0.5$.
If the condition (\ref{eq:groomingtest}) is true, then stop and return 
all particles that remain in the left hemisphere. If it is false, 
remove the lower energy branch, and continue to the next branching at smaller angle. 
Repeat the procedure for the other hemisphere.
\item Once the groomer has terminated, any observable can be measured on 
the particles that remain in the two hemispheres.
\end{enumerate}

In Ref.~\cite{Frye:2016aiz} a factorization theorem was derived for the cross 
section differential in the groomed hemisphere masses 
\begin{equation}
\tau_i = \frac{m_i^2}{E_i^2}
\,,\quad i=\textrm{L or R}
\end{equation}
for mass $m_i$ and energy $E_i$ of hemisphere $i$. For $\tau_i \ll\zcut \ll 1$, 
the cross section factorizes at all orders in perturbation theory as follows:
\begin{align}
\frac{1}{\sigma_0}\frac{\rd^2\sigma}{\rd\tau_\rL\, \rd\tau_\rR} = 
H(Q^2)S(\zcut) &\left[
J(\tau_\rL)\otimes S_c(\tau_\rL,\zcut)\right]  \\
\times&\left[J(\tau_\rR)\otimes S_c(\tau_\rR,\zcut)\right]\,,\nonumber
\end{align}
where $\sigma_0$ is the leading-order cross section for $e^+e^-\to q\bar q$,
$H(Q^2)$ is the hard function for quark--antiquark production in $e^+e^-$
collisions, $S(\zcut)$ is the global soft function for mMDT grooming, 
$J(\tau_i)$ is the quark jet function for hemisphere mass $\tau_i$, and $S_c(\tau_i,\zcut)$ is the collinear-soft function for hemisphere mass 
$\tau_i$ with mMDT grooming. The symbol $\otimes$ denotes convolution 
over the hemisphere mass $\tau_i$. In the functions we suppressed the
dependence on the renormalization scale $\mu$.  This factorization theorem is only valid in the limit in which $\zcut \ll 1$, and so corrections to it exist for any finite value of $\zcut$.  Recently, it was demonstrated that these finite $\zcut$ corrections to the factorization theorem are only at the level of a few percent in the $\tau_i \ll 1$ limit \cite{Larkoski:2020wgx}, so their fixed-order description is sufficient here.

Transforming into Laplace space, the cross section assumes a genuine factorized form,
\begin{align}\label{eq:laplacefact}
&\frac{\sigma(\nu_\rL,\nu_\rR)}{\sigma_0}\\
&
\hspace{0.5cm}=
H(Q^2)S(\zcut)\tilde J(\nu_L) 
\tilde S_c(\nu_\rL,\zcut)\tilde J(\nu_R) 
\tilde S_c(\nu_\rR,\zcut)\,,\nonumber
\end{align}
where $\nu_\rL$ ($\nu_\rR$) is the Laplace conjugate of $\tau_\rL$ ($\tau_\rR$).
In this product form, each function in the factorization theorem satisfies 
a simple renormalization group equation (RGE),
\begin{equation}
\mu\frac{\partial \tilde F}{\partial \mu} =
\left(d_F \Gamma_\cusp \log\frac{\mu^2}{\mu_F^2}
+\gamma_F\right) \tilde F
\,,\quad \tilde F = H
\,,\: S
\,,\: \tilde J
\,,\: \tilde S_c
\label{eq:RGE}
\end{equation}
where $d_F$ is a constant, $\mu_F$ is the canonical scale, and $\gamma_F$ 
is the non-cusp anomalous dimension, all depending on the function $\tilde F$.
$\Gamma_\cusp$ is the cusp anomalous dimension for back-to-back 
light-like Wilson lines in the fundamental representation of color SU(3).
Large logarithms of hemisphere masses can be resummed to all orders in \as\
using this renormalization group equation, whose exact solution is presented 
explicitly including ${\cal O}(\as^3)$ terms in Ref.~\cite{Abbate:2010xh}.
The order to which logarithms can be resummed using the RGE (\ref{eq:RGE}) 
depends on the accuracy to which its components are calculated. For the canonical definition of logarithmic accuracy \cite{Catani:1992ua}, \Tab{tab:logtab} 
shows the order in \as\ to which 
the components of the RGE are needed.  The two-loop soft function constants were calculated by the \textsc{SoftServe} collaboration \cite{Bell:2018vaa,Bell:2018oqa,Bell:2020yzz}.  In Ref.~\cite{Kardos:2020ppl} we extracted the last missing pieces 
needed for \NnLL{3} resummation of the distribution of jet masses with mMDT, namely
the two-loop constants $c_{S_c}^\mMDT$ of the collinear-soft function 
and the three-loop anomalous dimension of the global soft function 
$\gamma_S^\mMDT$ (in Laplace conjugate space),
\begin{align}\label{eq:twoloopconst}
c_{S_c}^\mMDT &=\left(\frac{\as}{4\pi}\right)^2 \left[ 
  C_F^2 \left(22\pm 4\right)
+ C_F C_A\left(41\pm 1\right)\right.\\
&
\hspace{3cm}\left.
+ C_F T_R n_f \left(14.4\pm 0.1\right)
\right]\,,\nonumber
\end{align}
with $C_F = 4/3$, $C_A = 3$, and $T_R=1/2$ in QCD, $n_f$ is the number of active quark flavors, and
\begin{equation}\label{eq:softanomdim}
\gamma_S^\mMDT = \left(\frac{\as}{4\pi}\right)^3
\left[-11600 \pm 2000\right] \qquad (n_f = 5)\,.
\end{equation}
These results enable resummation to \NnLL{3} accuracy for jet substructure 
observables that we present here for the first time.  As shown in Fig.~2 of Ref.~\cite{Kardos:2020ppl}, the extraction of the $C_F^2$ term of the two-loop constants of the collinear-soft function is especially numerically challenging, so a future direct calculation of these constants and anomalous dimensions would improve the quality and precision of the resummation results.

\begin{table}
\begin{center}
\begin{tabular}{c| c c c c c c}
order & \ \ $\Gamma_\cusp$\ \  & \ \ $\gamma_F$ \ \ &\ \  $\beta$ \ \ & \ \ $c_F$ \ \ &\ \  matching\\
 \hline
 $n=0$& \as & -  & \as & -  & - \\
 $n>0$ & $\as^{n+1}$ & $\as^n$ & $\as^{n+1}$ & $\as^{n-1}$ & $\as^n$\\
\end{tabular}
\end{center}
\caption{\label{tab:logtab}
\as-order of ingredients needed for resummation to the logarithmic accuracy given by logarithmic order \NnLL{n}. $\Gamma_\cusp$ is the cusp anomalous dimension, $\gamma_F$ is the non-cusp anomalous dimension for function $\tilde F$, $\beta$ is the QCD $\beta$-function, and $c_F$ are the low-scale constants for function $\tilde F$.  The final column shows the relative order to which the resummed cross section can be additively matched to fixed-order.}
\end{table}


We present predictions in perturbation theory for the single-differential 
cross section of the groomed heavy hemisphere mass $\frac{\rho}{\sigma_0}
\frac{\rd\sigma_g}{\rd\rho}$, defined as
\begin{align}
\frac{\rd\sigma_\rg}{\rd\rho} &= 
\int \!\rd\tau_\rL \, \rd\tau_\rR\, 
\frac{\rd^2\sigma}{\rd\tau_\rL\, \rd\tau_\rR}\left[
\Theta(\tau_\rL - \tau_\rR)\,\delta(\rho - \tau_\rL)\right.\\
&\hspace{3.5cm}\left.+
\Theta(\tau_\rR - \tau_\rL)\,\delta(\rho - \tau_\rR)
\right]\,,\nonumber
\end{align}
where the subscript g on the cross section indicates that it is groomed. 
This definition of the heavy hemisphere mass differs from the standard 
definition of the ungroomed case when the heavy hemisphere mass is defined as:
$\rho = \frac{\max(m_\rL^2,m_\rR^2)}{Q^2}$,
with $Q$ being the center-of-mass energy.  When hemispheres are groomed, 
the grooming eliminates their dominant correlations, and so it is more 
natural to define the groomed mass with respect to the hemisphere energy, 
and not the center-of-mass energy.

\begin{figure}[t!]
\centering
\includegraphics[width=0.6\linewidth]{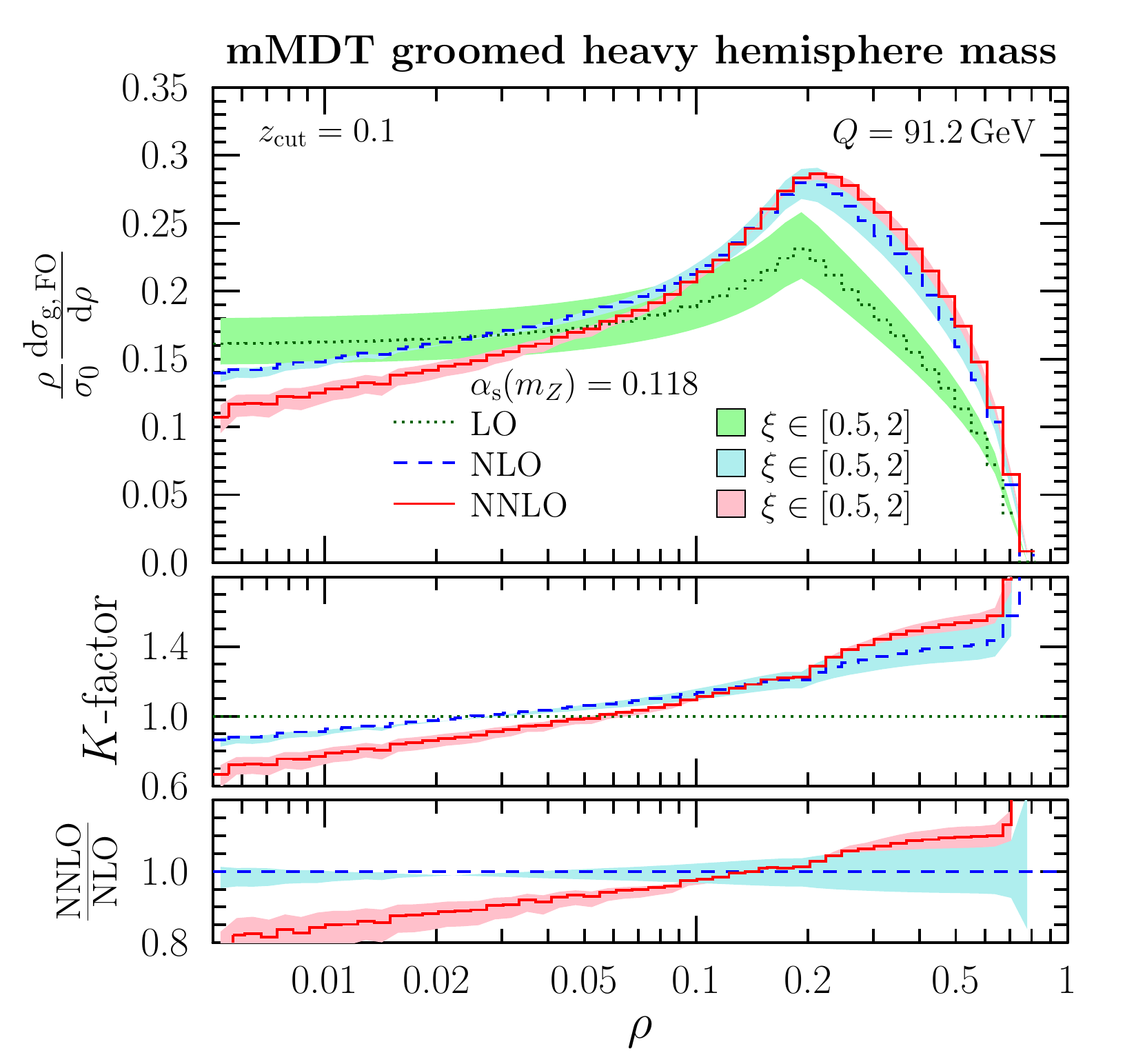}\\
\vspace{0.2cm}
\includegraphics[width=0.6\linewidth]{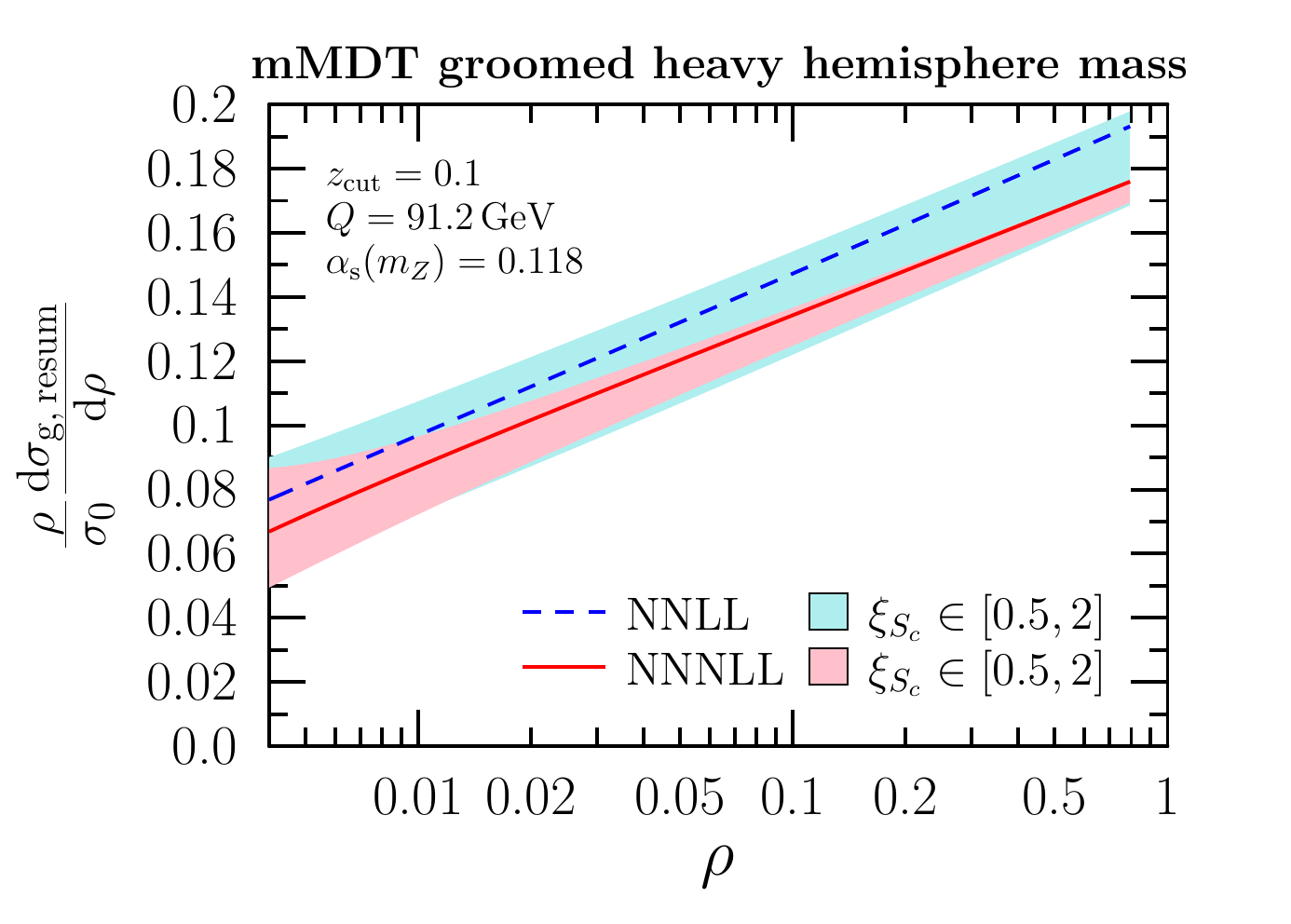}
\caption{Predictions for the groomed heavy jet mass in perturbation theory with $\zcut=0.1$. 
Top: at LO, NLO and NNLO accuracies, and their ratios.  The bands represent the uncertainties due to the variation of the renormalization 
scale $\mu = \xi Q$ in the range $\xi \in [1/2,2]$. Bottom: \NnLL{2} and \NnLL{3} accurate distributions. The bands represent the uncertainties due to the variation of the collinear-soft scale
$\mu_{S_c} = \xi_{S_c} 2e^{-\gamma_E} \sqrt{\zcut\rho}\,Q$ in the range $\xi_{S_c} \in [1/2,2]$.}
\label{fig:FO-resum}
\end{figure}

The CoLoRFulNNLO subtraction method was developed to compute QCD jet 
cross sections at the \NNLO\ accuracy. Currently it is completed for processes 
without colored particles in the initial states, and it is implemented in 
the \MCCSM\ code (Monte Carlo for the CoLoRFulNNLO Subtraction Method) \cite{DelDuca:2016csb,DelDuca:2016ily,Tulipant:2017ybb,Kardos:2016pic,Kardos:2018kth}.  This program can be used to compute the differential cross section of the 
mMDT groomed heavy hemisphere mass at fixed order in perturbation theory. 
\MCCSM\ calculates directly the $\rho$-dependent coefficients $A$, $B$, and $C$ (times their respective coupling factors) in the differential distribution
\begin{align}\label{eq:nnlo}
\rho &\frac{\rd\sigma_{\rg,{\rm NNLO}}}{\rd\rho} = \frac{\alpha_s}{2\pi}A_\rg+\left(
\frac{\alpha_s}{2\pi}
\right)^2\left[
B_\rg+A_\rg\beta_0 \ln\xi
\right]\\
&
+\left(
\frac{\alpha_s}{2\pi}
\right)^3\left[
C_\rg+2B_\rg\beta_0 \ln\xi\right.
\left.+A_\rg\left(\frac{\beta_1}{2} \ln\xi+\beta_0^2\ln^2\xi
\right)
\right]\nonumber,
\end{align}
where $\as = \as(\mu)$ is the strong coupling evaluated at the renormalization 
scale $\mu=\xi Q$, $\beta_0$ and $\beta_1$ are the first two coefficients in the 
perturbative expansion of the QCD $\beta$-function and $Q$ is the 
center-of-mass collision energy. 
We present the predictions of \MCCSM\ for the normalized 
cross section $\frac{\rho}{\sigma_0}
\frac{\rd\sigma_{\rg}}{\rd\rho}$ 
at the first three orders in perturbation theory (LO, NLO and \NNLO) in the top panel of 
\fig{fig:FO-resum}. The lower panels exhibit the K-factors defined as
\begin{equation}
K_{\rm FO/LO}(\xi)=\frac{\left(\rd\sigma_{\rm g,FO}(\mu=\xi Q)/\rd \rho\right)}{\left(\rd\sigma_{\rm g,LO}(\mu=Q)/\rd \rho\right)}
\,,
\end{equation}
and the ratio $K_{\rm NNLO/NLO}$.
We see that the ${\cal O}(\as^3)$ corrections stabilize the dependence on the renormalization scale for large values of $\rho$ ($\rho > 0.1$) as expected, while the predictions are clearly not reliable for $\rho\ll 0.1$. To stabilize the latter we need to resum the large logarithmic contributions.

All functions that appear in the factorization formula Eq.~\ref{eq:laplacefact} can also be found explicitly in Ref.~\cite{Frye:2016aiz}, including their matrix-element definitions.  Due to the factorized form of the cross section, each function in the factorization theorem has its own natural scale at which it is defined, and they can be varied independently to provide some estimate of residual scale uncertainties.  We leave a detailed scale variation study to future work, and here we just vary the scale of the collinear-soft function 
$\mu_{S_c}= \xi_{S_c} 2e^{-\gamma_E} \sqrt{\zcut\rho}\,Q$ in the range $\xi_{S_c} \in [1/2,2]$.  
The collinear-soft function is the lowest scale function in the factorization theorem, so variations of its scale will at least be representative of a more complete analysis.  Additionally, we just use the central values of the two-loop constant and three-loop anomalous dimension of Eqs.~\ref{eq:twoloopconst} and \ref{eq:softanomdim}, with no inclusion of their uncertainty.  We present the resummed predictions at \NnLL{2} and \NnLL{3} accuracies for the normalized cross section 
$\frac{\rho}{\sigma_0}
\frac{\rd\sigma_{\rg}}{\rd\rho}$ in the bottom panel of \fig{fig:FO-resum}. We see that these predictions are stable against the variation of the collinear-soft scale, but the range of validity is confined to $\rho \ll \zcut \ll 1$.

\begin{figure}[t!]
\centering
\includegraphics[width=0.6\linewidth]{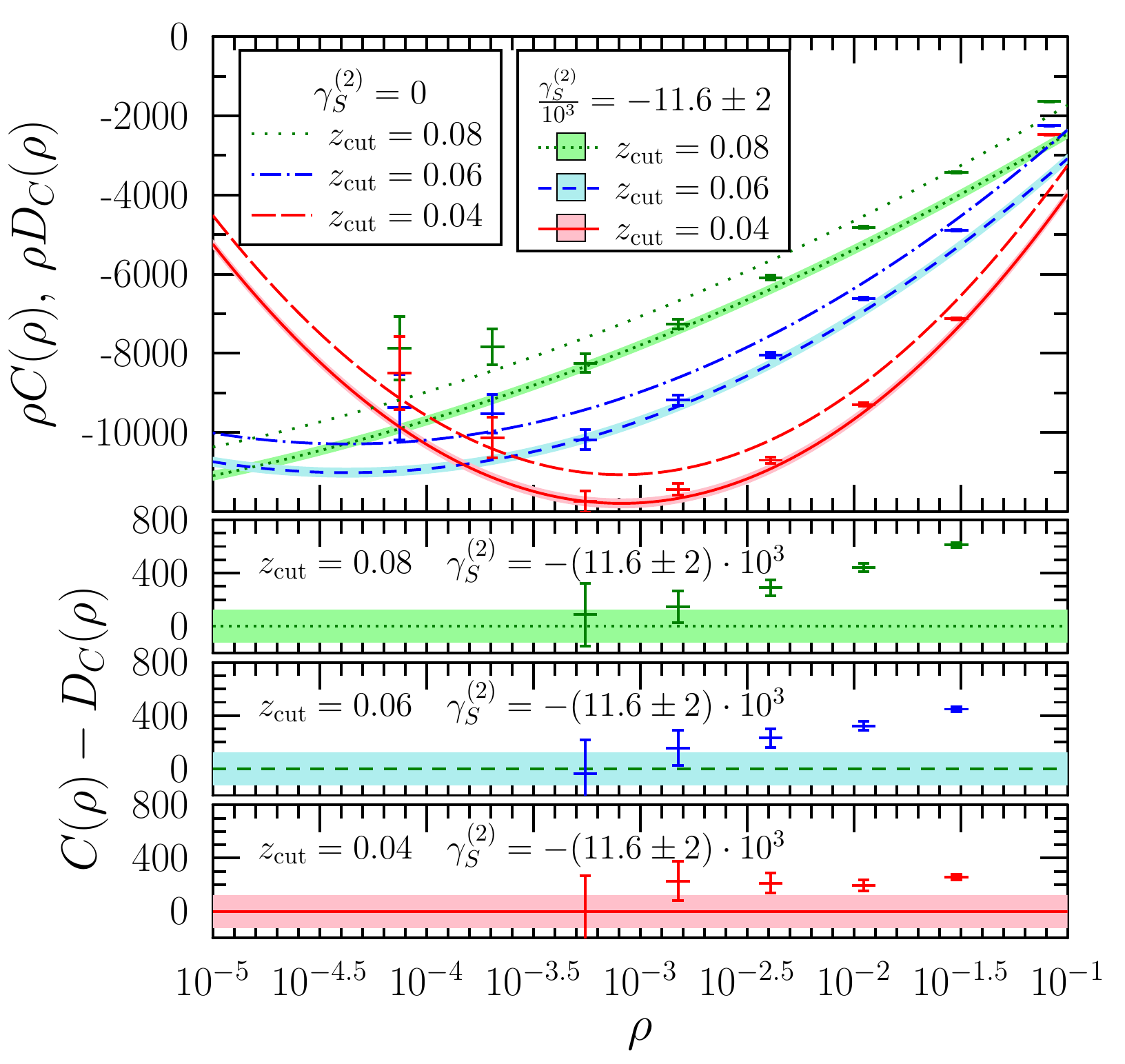}\\
\vspace{0.2cm}
\includegraphics[width=0.6\linewidth]{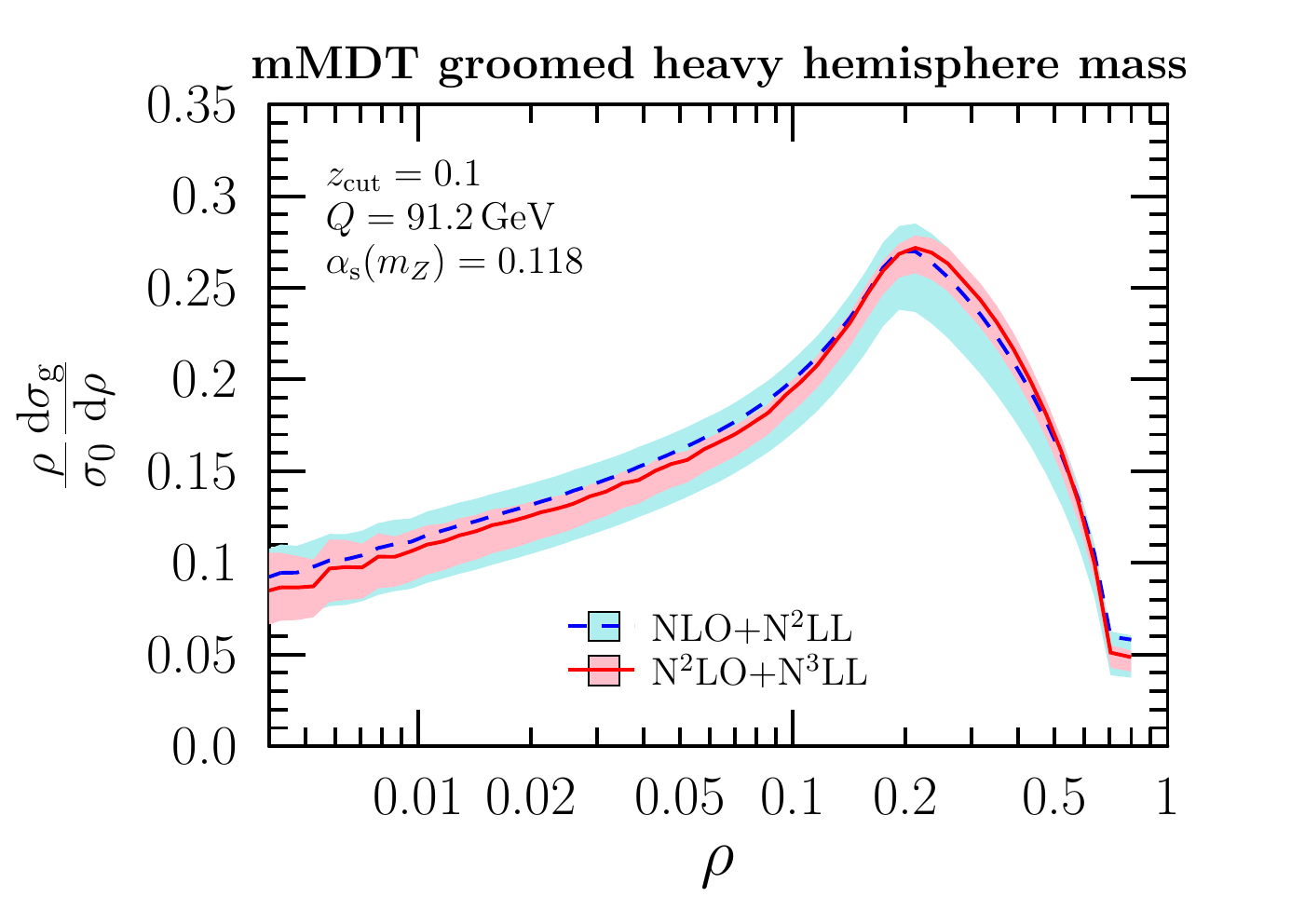}
\caption{Predictions for the groomed heavy jet mass in perturbation theory. 
Top: Comparison of the ${\cal O}(\as^3)$ coefficients at full fixed order and 
at leading power in $\rho$. Bottom: Predictions at matched NLO+\NnLL{2} and \NNLO+\NnLL{3} accuracy with $\zcut = 0.1$. 
The bands represent the uncertainties due to the variation of the renormalization and collinear-soft 
scales in the range [1/2,2] times their respective default scales.}
\label{fig:matching}
\end{figure}

The regions of validity of the predictions at \NNLO\ and at \NnLL{3} are 
complementary, the former gives a good description for large, 
while the latter for small values of $\rho$. In order to extend the precise 
description over the full phase space, the fixed-order and resummed 
predictions have to be matched. The additive matching requires the elimination 
of the logarithmic terms that are present in both predictions. The coefficients 
in the expansion of the resummed prediction in \as,
\begin{align}
\frac{\rd\sigma_\text{g,LP}}{\rd\rho} = \delta(\rho)D_{\delta,\rg}
&+ \frac{\as}{2\pi}(D_{A,\rg}(\rho))_+
+ \left(\frac{\as}{2\pi}\right)^2(D_{B,\rg}(\rho))_+\nonumber\\
&
+ \left(\frac{\as}{2\pi}\right)^3(D_{C,\rg}(\rho))_+
\,,
\end{align}
can be found in Ref.~\cite{Kardos:2020ppl} including the ${\cal O}(\as^3)$ coefficient. For $\rho > 0$ the 
$\delta$-functions can be ignored and $+$-distributions reduce to simple functions of $\rho$.
We compare the $D_{C,\rg}(\rho)$ function to the $C_\rg(\rho)$ coefficient in the 
fixed-order expansion in the top panel of \fig{fig:matching} where we show the logarithmic expansion with two assumed values of the three-loop non-cusp anomalous dimension $\gamma_S^{(2)}$: 0 and our extracted value with uncertainties from Eq.~\ref{eq:softanomdim}.  As the value of $\zcut$ is decreased, improved agreement between the {\tt MCCSM} results and the singular distribution is observed at small $\rho$, down to about $\rho \sim 10^{-4}$ where numerical instabilities in {\tt MCCSM} become significant.

Subtracting this singular distribution from the sum of the \NNLO\ and \NnLL{3}, we obtain a prediction in perturbation theory with highest available accuracy:
\begin{align}
&\frac{\rho}{\sigma_0}
\frac{\rd\sigma_{\rg}}{\rd\rho}
= \frac{\rho}{\sigma_0} \left(
\frac{\rd\sigma_{\rg,{\rm N^3LL}}}{\rd\rho}
+ \frac{\rd\sigma_{\rg,{\rm N^2LO}}}{\rd\rho}
- \frac{\rd\sigma_{\rg,{\rm LP}}}{\rd\rho}
\right)
\,,
\end{align}
which we present in the bottom panel of \fig{fig:matching}. Good convergence of the matched predictions is observed for all values of $\rho$, with the results at \NNLO+\NnLL{3} lying within the scale variation bands of the NLO+\NnLL{2} prediction.  We have truncated this perturbative prediction at a value of $\rho$ that lies above the region in which non-perturbative physics dominates the distribution.  


We have demonstrated the highest precision perturbative predictions for groomed jets in $e^+e^-$ collisions.  These results are sufficiently accurate to enable extraction of $\alpha_s$, when combined with leading corrections due to non-perturbative physics.  While there is no currently-running $e^+e^-$ collider, analyses of archived LEP data have been completed \cite{Badea:2019vey}, and the results presented here motivate further measurements on these archived data.  Due to the elimination of soft radiation with mMDT grooming, the collinear-soft and jet functions in the factorization theorem are identical to that for corresponding measurements at hadron colliders.  Thus, we anticipate these results can be used to further improve the theory-data comparisons of groomed jet masses measured at ATLAS and CMS \cite{Aaboud:2017qwh,Sirunyan:2018xdh,Aad:2019vyi}, and, along with continual advances in fixed-order predictions, enable precision extractions of fundamental constants at the LHC.

 We thank Guido Bell, Jim Talbert, and Rudi Rahn for providing their results for the two-loop constants of the global soft function.  This work was facilitated in part by the Portland Institute for Computational Science and its resources acquired using NSF Grant DMS 1624776.
This work was supported by grant K 125105 of the National Research, Development and Innovation Fund in Hungary and the Premium Postdoctoral Fellowship program of the Hungarian Academy of Sciences.

\bibliography{SoftDropNNNLL}

\providecommand{\noopsort}[1]{}\providecommand{\singleletter}[1]{#1}%
\begin{thebibliography}{44}%
\makeatletter
\providecommand \@ifxundefined [1]{%
 \@ifx{#1\undefined}
}%
\providecommand \@ifnum [1]{%
 \ifnum #1\expandafter \@firstoftwo
 \else \expandafter \@secondoftwo
 \fi
}%
\providecommand \@ifx [1]{%
 \ifx #1\expandafter \@firstoftwo
 \else \expandafter \@secondoftwo
 \fi
}%
\providecommand \natexlab [1]{#1}%
\providecommand \enquote  [1]{``#1''}%
\providecommand \bibnamefont  [1]{#1}%
\providecommand \bibfnamefont [1]{#1}%
\providecommand \citenamefont [1]{#1}%
\providecommand \href@noop [0]{\@secondoftwo}%
\providecommand \href [0]{\begingroup \@sanitize@url \@href}%
\providecommand \@href[1]{\@@startlink{#1}\@@href}%
\providecommand \@@href[1]{\endgroup#1\@@endlink}%
\providecommand \@sanitize@url [0]{\catcode `\\12\catcode `\$12\catcode
  `\&12\catcode `\#12\catcode `\^12\catcode `\_12\catcode `\%12\relax}%
\providecommand \@@startlink[1]{}%
\providecommand \@@endlink[0]{}%
\providecommand \url  [0]{\begingroup\@sanitize@url \@url }%
\providecommand \@url [1]{\endgroup\@href {#1}{\urlprefix }}%
\providecommand \urlprefix  [0]{URL }%
\providecommand \Eprint [0]{\href }%
\providecommand \doibase [0]{http://dx.doi.org/}%
\providecommand \selectlanguage [0]{\@gobble}%
\providecommand \bibinfo  [0]{\@secondoftwo}%
\providecommand \bibfield  [0]{\@secondoftwo}%
\providecommand \translation [1]{[#1]}%
\providecommand \BibitemOpen [0]{}%
\providecommand \bibitemStop [0]{}%
\providecommand \bibitemNoStop [0]{.\EOS\space}%
\providecommand \EOS [0]{\spacefactor3000\relax}%
\providecommand \BibitemShut  [1]{\csname bibitem#1\endcsname}%
\let\auto@bib@innerbib\@empty
\bibitem [{\citenamefont {Decamp}\ \emph {et~al.}(1991)\citenamefont {Decamp}
  \emph {et~al.}}]{Decamp:1990cg}%
  \BibitemOpen
  \bibfield  {author} {\bibinfo {author} {\bibfnamefont {D.}~\bibnamefont
  {Decamp}} \emph {et~al.} (\bibinfo {collaboration} {ALEPH}),\ }\href
  {\doibase 10.1016/0370-2693(91)90278-X} {\bibfield  {journal} {\bibinfo
  {journal} {Phys. Lett.}\ }\textbf {\bibinfo {volume} {B255}},\ \bibinfo
  {pages} {623} (\bibinfo {year} {1991})}\BibitemShut {NoStop}%
\bibitem [{\citenamefont {Adeva}\ \emph {et~al.}(1990)\citenamefont {Adeva}
  \emph {et~al.}}]{Adeva:1990nu}%
  \BibitemOpen
  \bibfield  {author} {\bibinfo {author} {\bibfnamefont {B.}~\bibnamefont
  {Adeva}} \emph {et~al.} (\bibinfo {collaboration} {L3}),\ }\href {\doibase
  10.1016/0370-2693(90)90323-X} {\bibfield  {journal} {\bibinfo  {journal}
  {Phys. Lett.}\ }\textbf {\bibinfo {volume} {B248}},\ \bibinfo {pages} {464}
  (\bibinfo {year} {1990})}\BibitemShut {NoStop}%
\bibitem [{\citenamefont {Adeva}\ \emph {et~al.}(1991)\citenamefont {Adeva}
  \emph {et~al.}}]{Adeva:1991nv}%
  \BibitemOpen
  \bibfield  {author} {\bibinfo {author} {\bibfnamefont {B.}~\bibnamefont
  {Adeva}} \emph {et~al.} (\bibinfo {collaboration} {L3}),\ }\href {\doibase
  10.1016/0370-2693(91)90118-A} {\bibfield  {journal} {\bibinfo  {journal}
  {Phys. Lett.}\ }\textbf {\bibinfo {volume} {B271}},\ \bibinfo {pages} {461}
  (\bibinfo {year} {1991})}\BibitemShut {NoStop}%
\bibitem [{\citenamefont {Abreu}\ \emph {et~al.}(1992)\citenamefont {Abreu}
  \emph {et~al.}}]{Abreu:1992yc}%
  \BibitemOpen
  \bibfield  {author} {\bibinfo {author} {\bibfnamefont {P.}~\bibnamefont
  {Abreu}} \emph {et~al.} (\bibinfo {collaboration} {DELPHI}),\ }\href
  {\doibase 10.1007/BF01881708} {\bibfield  {journal} {\bibinfo  {journal} {Z.
  Phys.}\ }\textbf {\bibinfo {volume} {C54}},\ \bibinfo {pages} {55} (\bibinfo
  {year} {1992})}\BibitemShut {NoStop}%
\bibitem [{\citenamefont {Adrian}\ \emph {et~al.}(1992)\citenamefont {Adrian}
  \emph {et~al.}}]{Adriani:1992gs}%
  \BibitemOpen
  \bibfield  {author} {\bibinfo {author} {\bibfnamefont {O.}~\bibnamefont
  {Adrian}} \emph {et~al.} (\bibinfo {collaboration} {L3}),\ }\href {\doibase
  10.1016/0370-2693(92)90463-E} {\bibfield  {journal} {\bibinfo  {journal}
  {Phys. Lett.}\ }\textbf {\bibinfo {volume} {B284}},\ \bibinfo {pages} {471}
  (\bibinfo {year} {1992})}\BibitemShut {NoStop}%
\bibitem [{\citenamefont {Decamp}\ \emph {et~al.}(1992)\citenamefont {Decamp}
  \emph {et~al.}}]{Decamp:1992wz}%
  \BibitemOpen
  \bibfield  {author} {\bibinfo {author} {\bibfnamefont {D.}~\bibnamefont
  {Decamp}} \emph {et~al.} (\bibinfo {collaboration} {ALEPH}),\ }\href
  {\doibase 10.1016/0370-2693(92)91942-3} {\bibfield  {journal} {\bibinfo
  {journal} {Phys. Lett.}\ }\textbf {\bibinfo {volume} {B284}},\ \bibinfo
  {pages} {163} (\bibinfo {year} {1992})}\BibitemShut {NoStop}%
\bibitem [{\citenamefont {Acton}\ \emph {et~al.}(1993)\citenamefont {Acton}
  \emph {et~al.}}]{Acton:1993zh}%
  \BibitemOpen
  \bibfield  {author} {\bibinfo {author} {\bibfnamefont {P.~D.}\ \bibnamefont
  {Acton}} \emph {et~al.} (\bibinfo {collaboration} {OPAL}),\ }\href {\doibase
  10.1007/BF01555834} {\bibfield  {journal} {\bibinfo  {journal} {Z. Phys.}\
  }\textbf {\bibinfo {volume} {C59}},\ \bibinfo {pages} {1} (\bibinfo {year}
  {1993})}\BibitemShut {NoStop}%
\bibitem [{\citenamefont {Abreu}\ \emph {et~al.}(1999)\citenamefont {Abreu}
  \emph {et~al.}}]{Abreu:1999rc}%
  \BibitemOpen
  \bibfield  {author} {\bibinfo {author} {\bibfnamefont {P.}~\bibnamefont
  {Abreu}} \emph {et~al.} (\bibinfo {collaboration} {DELPHI}),\ }\href
  {\doibase 10.1016/S0370-2693(99)00472-4} {\bibfield  {journal} {\bibinfo
  {journal} {Phys. Lett.}\ }\textbf {\bibinfo {volume} {B456}},\ \bibinfo
  {pages} {322} (\bibinfo {year} {1999})}\BibitemShut {NoStop}%
\bibitem [{\citenamefont {Abbiendi}\ \emph {et~al.}(2001)\citenamefont
  {Abbiendi} \emph {et~al.}}]{Abbiendi:2001qn}%
  \BibitemOpen
  \bibfield  {author} {\bibinfo {author} {\bibfnamefont {G.}~\bibnamefont
  {Abbiendi}} \emph {et~al.} (\bibinfo {collaboration} {OPAL}),\ }\href
  {\doibase 10.1007/s100520100699} {\bibfield  {journal} {\bibinfo  {journal}
  {Eur. Phys. J.}\ }\textbf {\bibinfo {volume} {C20}},\ \bibinfo {pages} {601}
  (\bibinfo {year} {2001})},\ \Eprint {http://arxiv.org/abs/hep-ex/0101044}
  {arXiv:hep-ex/0101044 [hep-ex]} \BibitemShut {NoStop}%
\bibitem [{\citenamefont {Heister}\ \emph {et~al.}(2003)\citenamefont {Heister}
  \emph {et~al.}}]{Heister:2002tq}%
  \BibitemOpen
  \bibfield  {author} {\bibinfo {author} {\bibfnamefont {A.}~\bibnamefont
  {Heister}} \emph {et~al.} (\bibinfo {collaboration} {ALEPH}),\ }\href
  {\doibase 10.1140/epjc/s2002-01114-2} {\bibfield  {journal} {\bibinfo
  {journal} {Eur. Phys. J.}\ }\textbf {\bibinfo {volume} {C27}},\ \bibinfo
  {pages} {1} (\bibinfo {year} {2003})}\BibitemShut {NoStop}%
\bibitem [{\citenamefont {Abbiendi}\ \emph {et~al.}(2005)\citenamefont
  {Abbiendi} \emph {et~al.}}]{Abbiendi:2004qz}%
  \BibitemOpen
  \bibfield  {author} {\bibinfo {author} {\bibfnamefont {G.}~\bibnamefont
  {Abbiendi}} \emph {et~al.} (\bibinfo {collaboration} {OPAL}),\ }\href
  {\doibase 10.1140/epjc/s2005-02120-6} {\bibfield  {journal} {\bibinfo
  {journal} {Eur. Phys. J.}\ }\textbf {\bibinfo {volume} {C40}},\ \bibinfo
  {pages} {287} (\bibinfo {year} {2005})},\ \Eprint
  {http://arxiv.org/abs/hep-ex/0503051} {arXiv:hep-ex/0503051 [hep-ex]}
  \BibitemShut {NoStop}%
\bibitem [{\citenamefont {Tanabashi}\ \emph {et~al.}(2018)\citenamefont
  {Tanabashi} \emph {et~al.}}]{Tanabashi:2018oca}%
  \BibitemOpen
  \bibfield  {author} {\bibinfo {author} {\bibfnamefont {M.}~\bibnamefont
  {Tanabashi}} \emph {et~al.} (\bibinfo {collaboration} {Particle Data
  Group}),\ }\href {\doibase 10.1103/PhysRevD.98.030001} {\bibfield  {journal}
  {\bibinfo  {journal} {Phys. Rev.}\ }\textbf {\bibinfo {volume} {D98}},\
  \bibinfo {pages} {030001} (\bibinfo {year} {2018})}\BibitemShut {NoStop}%
\bibitem [{\citenamefont {Salam}(2019)}]{Salam:2017qdl}%
  \BibitemOpen
  \bibfield  {author} {\bibinfo {author} {\bibfnamefont {G.~P.}\ \bibnamefont
  {Salam}},\ }in\ \href {\doibase 10.1142/9789813238053_0007} {\emph {\bibinfo
  {booktitle} {From My Vast Repertoire ...: Guido Altarelli's Legacy}}},\
  \bibinfo {editor} {edited by\ \bibinfo {editor} {\bibfnamefont
  {A.}~\bibnamefont {Levy}}, \bibinfo {editor} {\bibfnamefont {S.}~\bibnamefont
  {Forte}}, \ and\ \bibinfo {editor} {\bibfnamefont {G.}~\bibnamefont
  {Ridolfi}}}\ (\bibinfo {year} {2019})\ pp.\ \bibinfo {pages} {101--121},\
  \Eprint {http://arxiv.org/abs/1712.05165} {arXiv:1712.05165 [hep-ph]}
  \BibitemShut {NoStop}%
\bibitem [{\citenamefont {Tr\'ocs\'anyi}(2019)}]{Trocsanyi:2019irv}%
  \BibitemOpen
  \bibfield  {author} {\bibinfo {author} {\bibfnamefont {Z.~L.}\ \bibnamefont
  {Tr\'ocs\'anyi}},\ }\bibfield  {booktitle} {\emph {\bibinfo {booktitle}
  {{Proceedings, 18th Hellenic School and Workshops on Elementary Particle
  Physics and Gravity (CORFU2018): Corfu, Corfu, Greece}}},\ }\href {\doibase
  10.22323/1.347.0002} {\bibfield  {journal} {\bibinfo  {journal} {PoS}\
  }\textbf {\bibinfo {volume} {CORFU2018}},\ \bibinfo {pages} {002} (\bibinfo
  {year} {2019})}\BibitemShut {NoStop}%
\bibitem [{\citenamefont {Dokshitzer}\ \emph {et~al.}(1996)\citenamefont
  {Dokshitzer}, \citenamefont {Marchesini},\ and\ \citenamefont
  {Webber}}]{Dokshitzer:1995qm}%
  \BibitemOpen
  \bibfield  {author} {\bibinfo {author} {\bibfnamefont {Y.~L.}\ \bibnamefont
  {Dokshitzer}}, \bibinfo {author} {\bibfnamefont {G.}~\bibnamefont
  {Marchesini}}, \ and\ \bibinfo {author} {\bibfnamefont {B.~R.}\ \bibnamefont
  {Webber}},\ }\href {\doibase 10.1016/0550-3213(96)00155-1} {\bibfield
  {journal} {\bibinfo  {journal} {Nucl. Phys.}\ }\textbf {\bibinfo {volume}
  {B469}},\ \bibinfo {pages} {93} (\bibinfo {year} {1996})},\ \Eprint
  {http://arxiv.org/abs/hep-ph/9512336} {arXiv:hep-ph/9512336 [hep-ph]}
  \BibitemShut {NoStop}%
\bibitem [{\citenamefont {Verbytskyi}\ \emph {et~al.}(2019)\citenamefont
  {Verbytskyi}, \citenamefont {Banfi}, \citenamefont {Kardos}, \citenamefont
  {Monni}, \citenamefont {Kluth}, \citenamefont {Somogyi}, \citenamefont
  {Sz\H{o}r}, \citenamefont {Tr\'ocs\'anyi}, \citenamefont {Tulip\'ant},\ and\
  \citenamefont {Zanderighi}}]{Verbytskyi:2019zhh}%
  \BibitemOpen
  \bibfield  {author} {\bibinfo {author} {\bibfnamefont {A.}~\bibnamefont
  {Verbytskyi}}, \bibinfo {author} {\bibfnamefont {A.}~\bibnamefont {Banfi}},
  \bibinfo {author} {\bibfnamefont {A.}~\bibnamefont {Kardos}}, \bibinfo
  {author} {\bibfnamefont {P.~F.}\ \bibnamefont {Monni}}, \bibinfo {author}
  {\bibfnamefont {S.}~\bibnamefont {Kluth}}, \bibinfo {author} {\bibfnamefont
  {G.}~\bibnamefont {Somogyi}}, \bibinfo {author} {\bibfnamefont
  {Z.}~\bibnamefont {Sz\H{o}r}}, \bibinfo {author} {\bibfnamefont
  {Z.}~\bibnamefont {Tr\'ocs\'anyi}}, \bibinfo {author} {\bibfnamefont
  {Z.}~\bibnamefont {Tulip\'ant}}, \ and\ \bibinfo {author} {\bibfnamefont
  {G.}~\bibnamefont {Zanderighi}},\ }\href {\doibase 10.1007/JHEP08(2019)129}
  {\bibfield  {journal} {\bibinfo  {journal} {JHEP}\ }\textbf {\bibinfo
  {volume} {08}},\ \bibinfo {pages} {129} (\bibinfo {year} {2019})},\ \Eprint
  {http://arxiv.org/abs/1902.08158} {arXiv:1902.08158 [hep-ph]} \BibitemShut
  {NoStop}%
\bibitem [{\citenamefont {Becher}\ and\ \citenamefont
  {Schwartz}(2008)}]{Becher:2008cf}%
  \BibitemOpen
  \bibfield  {author} {\bibinfo {author} {\bibfnamefont {T.}~\bibnamefont
  {Becher}}\ and\ \bibinfo {author} {\bibfnamefont {M.~D.}\ \bibnamefont
  {Schwartz}},\ }\href {\doibase 10.1088/1126-6708/2008/07/034} {\bibfield
  {journal} {\bibinfo  {journal} {JHEP}\ }\textbf {\bibinfo {volume} {07}},\
  \bibinfo {pages} {034} (\bibinfo {year} {2008})},\ \Eprint
  {http://arxiv.org/abs/0803.0342} {arXiv:0803.0342 [hep-ph]} \BibitemShut
  {NoStop}%
\bibitem [{\citenamefont {Hoang}\ \emph
  {et~al.}(2015{\natexlab{a}})\citenamefont {Hoang}, \citenamefont
  {Kolodrubetz}, \citenamefont {Mateu},\ and\ \citenamefont
  {Stewart}}]{Hoang:2014wka}%
  \BibitemOpen
  \bibfield  {author} {\bibinfo {author} {\bibfnamefont {A.~H.}\ \bibnamefont
  {Hoang}}, \bibinfo {author} {\bibfnamefont {D.~W.}\ \bibnamefont
  {Kolodrubetz}}, \bibinfo {author} {\bibfnamefont {V.}~\bibnamefont {Mateu}},
  \ and\ \bibinfo {author} {\bibfnamefont {I.~W.}\ \bibnamefont {Stewart}},\
  }\href {\doibase 10.1103/PhysRevD.91.094017} {\bibfield  {journal} {\bibinfo
  {journal} {Phys. Rev.}\ }\textbf {\bibinfo {volume} {D91}},\ \bibinfo {pages}
  {094017} (\bibinfo {year} {2015}{\natexlab{a}})},\ \Eprint
  {http://arxiv.org/abs/1411.6633} {arXiv:1411.6633 [hep-ph]} \BibitemShut
  {NoStop}%
\bibitem [{\citenamefont {Abbate}\ \emph {et~al.}(2011)\citenamefont {Abbate},
  \citenamefont {Fickinger}, \citenamefont {Hoang}, \citenamefont {Mateu},\
  and\ \citenamefont {Stewart}}]{Abbate:2010xh}%
  \BibitemOpen
  \bibfield  {author} {\bibinfo {author} {\bibfnamefont {R.}~\bibnamefont
  {Abbate}}, \bibinfo {author} {\bibfnamefont {M.}~\bibnamefont {Fickinger}},
  \bibinfo {author} {\bibfnamefont {A.~H.}\ \bibnamefont {Hoang}}, \bibinfo
  {author} {\bibfnamefont {V.}~\bibnamefont {Mateu}}, \ and\ \bibinfo {author}
  {\bibfnamefont {I.~W.}\ \bibnamefont {Stewart}},\ }\href {\doibase
  10.1103/PhysRevD.83.074021} {\bibfield  {journal} {\bibinfo  {journal} {Phys.
  Rev.}\ }\textbf {\bibinfo {volume} {D83}},\ \bibinfo {pages} {074021}
  (\bibinfo {year} {2011})},\ \Eprint {http://arxiv.org/abs/1006.3080}
  {arXiv:1006.3080 [hep-ph]} \BibitemShut {NoStop}%
\bibitem [{\citenamefont {Hoang}\ \emph
  {et~al.}(2015{\natexlab{b}})\citenamefont {Hoang}, \citenamefont
  {Kolodrubetz}, \citenamefont {Mateu},\ and\ \citenamefont
  {Stewart}}]{Hoang:2015hka}%
  \BibitemOpen
  \bibfield  {author} {\bibinfo {author} {\bibfnamefont {A.~H.}\ \bibnamefont
  {Hoang}}, \bibinfo {author} {\bibfnamefont {D.~W.}\ \bibnamefont
  {Kolodrubetz}}, \bibinfo {author} {\bibfnamefont {V.}~\bibnamefont {Mateu}},
  \ and\ \bibinfo {author} {\bibfnamefont {I.~W.}\ \bibnamefont {Stewart}},\
  }\href {\doibase 10.1103/PhysRevD.91.094018} {\bibfield  {journal} {\bibinfo
  {journal} {Phys. Rev.}\ }\textbf {\bibinfo {volume} {D91}},\ \bibinfo {pages}
  {094018} (\bibinfo {year} {2015}{\natexlab{b}})},\ \Eprint
  {http://arxiv.org/abs/1501.04111} {arXiv:1501.04111 [hep-ph]} \BibitemShut
  {NoStop}%
\bibitem [{\citenamefont {Dasgupta}\ \emph
  {et~al.}(2013{\natexlab{a}})\citenamefont {Dasgupta}, \citenamefont
  {Fregoso}, \citenamefont {Marzani},\ and\ \citenamefont
  {Salam}}]{Dasgupta:2013ihk}%
  \BibitemOpen
  \bibfield  {author} {\bibinfo {author} {\bibfnamefont {M.}~\bibnamefont
  {Dasgupta}}, \bibinfo {author} {\bibfnamefont {A.}~\bibnamefont {Fregoso}},
  \bibinfo {author} {\bibfnamefont {S.}~\bibnamefont {Marzani}}, \ and\
  \bibinfo {author} {\bibfnamefont {G.~P.}\ \bibnamefont {Salam}},\ }\href
  {\doibase 10.1007/JHEP09(2013)029} {\bibfield  {journal} {\bibinfo  {journal}
  {JHEP}\ }\textbf {\bibinfo {volume} {09}},\ \bibinfo {pages} {029} (\bibinfo
  {year} {2013}{\natexlab{a}})},\ \Eprint {http://arxiv.org/abs/1307.0007}
  {arXiv:1307.0007 [hep-ph]} \BibitemShut {NoStop}%
\bibitem [{\citenamefont {Dasgupta}\ \emph
  {et~al.}(2013{\natexlab{b}})\citenamefont {Dasgupta}, \citenamefont
  {Fregoso}, \citenamefont {Marzani},\ and\ \citenamefont
  {Powling}}]{Dasgupta:2013via}%
  \BibitemOpen
  \bibfield  {author} {\bibinfo {author} {\bibfnamefont {M.}~\bibnamefont
  {Dasgupta}}, \bibinfo {author} {\bibfnamefont {A.}~\bibnamefont {Fregoso}},
  \bibinfo {author} {\bibfnamefont {S.}~\bibnamefont {Marzani}}, \ and\
  \bibinfo {author} {\bibfnamefont {A.}~\bibnamefont {Powling}},\ }\href
  {\doibase 10.1140/epjc/s10052-013-2623-3} {\bibfield  {journal} {\bibinfo
  {journal} {Eur. Phys. J.}\ }\textbf {\bibinfo {volume} {C73}},\ \bibinfo
  {pages} {2623} (\bibinfo {year} {2013}{\natexlab{b}})},\ \Eprint
  {http://arxiv.org/abs/1307.0013} {arXiv:1307.0013 [hep-ph]} \BibitemShut
  {NoStop}%
\bibitem [{\citenamefont {Larkoski}\ \emph {et~al.}(2014)\citenamefont
  {Larkoski}, \citenamefont {Marzani}, \citenamefont {Soyez},\ and\
  \citenamefont {Thaler}}]{Larkoski:2014wba}%
  \BibitemOpen
  \bibfield  {author} {\bibinfo {author} {\bibfnamefont {A.~J.}\ \bibnamefont
  {Larkoski}}, \bibinfo {author} {\bibfnamefont {S.}~\bibnamefont {Marzani}},
  \bibinfo {author} {\bibfnamefont {G.}~\bibnamefont {Soyez}}, \ and\ \bibinfo
  {author} {\bibfnamefont {J.}~\bibnamefont {Thaler}},\ }\href {\doibase
  10.1007/JHEP05(2014)146} {\bibfield  {journal} {\bibinfo  {journal} {JHEP}\
  }\textbf {\bibinfo {volume} {05}},\ \bibinfo {pages} {146} (\bibinfo {year}
  {2014})},\ \Eprint {http://arxiv.org/abs/1402.2657} {arXiv:1402.2657
  [hep-ph]} \BibitemShut {NoStop}%
\bibitem [{\citenamefont {Dasgupta}\ and\ \citenamefont
  {Salam}(2001)}]{Dasgupta:2001sh}%
  \BibitemOpen
  \bibfield  {author} {\bibinfo {author} {\bibfnamefont {M.}~\bibnamefont
  {Dasgupta}}\ and\ \bibinfo {author} {\bibfnamefont {G.~P.}\ \bibnamefont
  {Salam}},\ }\href {\doibase 10.1016/S0370-2693(01)00725-0} {\bibfield
  {journal} {\bibinfo  {journal} {Phys. Lett.}\ }\textbf {\bibinfo {volume}
  {B512}},\ \bibinfo {pages} {323} (\bibinfo {year} {2001})},\ \Eprint
  {http://arxiv.org/abs/hep-ph/0104277} {arXiv:hep-ph/0104277 [hep-ph]}
  \BibitemShut {NoStop}%
\bibitem [{\citenamefont {Baron}\ \emph {et~al.}(2018)\citenamefont {Baron},
  \citenamefont {Marzani},\ and\ \citenamefont {Theeuwes}}]{Baron:2018nfz}%
  \BibitemOpen
  \bibfield  {author} {\bibinfo {author} {\bibfnamefont {J.}~\bibnamefont
  {Baron}}, \bibinfo {author} {\bibfnamefont {S.}~\bibnamefont {Marzani}}, \
  and\ \bibinfo {author} {\bibfnamefont {V.}~\bibnamefont {Theeuwes}},\ }\href
  {\doibase 10.1007/JHEP08(2018)105, 10.1007/JHEP05(2019)056} {\bibfield
  {journal} {\bibinfo  {journal} {JHEP}\ }\textbf {\bibinfo {volume} {08}},\
  \bibinfo {pages} {105} (\bibinfo {year} {2018})},\ \bibinfo {note} {[erratum:
  JHEP05,056(2019)]},\ \Eprint {http://arxiv.org/abs/1803.04719}
  {arXiv:1803.04719 [hep-ph]} \BibitemShut {NoStop}%
\bibitem [{\citenamefont {Frye}\ \emph {et~al.}(2016)\citenamefont {Frye},
  \citenamefont {Larkoski}, \citenamefont {Schwartz},\ and\ \citenamefont
  {Yan}}]{Frye:2016aiz}%
  \BibitemOpen
  \bibfield  {author} {\bibinfo {author} {\bibfnamefont {C.}~\bibnamefont
  {Frye}}, \bibinfo {author} {\bibfnamefont {A.~J.}\ \bibnamefont {Larkoski}},
  \bibinfo {author} {\bibfnamefont {M.~D.}\ \bibnamefont {Schwartz}}, \ and\
  \bibinfo {author} {\bibfnamefont {K.}~\bibnamefont {Yan}},\ }\href {\doibase
  10.1007/JHEP07(2016)064} {\bibfield  {journal} {\bibinfo  {journal} {JHEP}\
  }\textbf {\bibinfo {volume} {07}},\ \bibinfo {pages} {064} (\bibinfo {year}
  {2016})},\ \Eprint {http://arxiv.org/abs/1603.09338} {arXiv:1603.09338
  [hep-ph]} \BibitemShut {NoStop}%
\bibitem [{\citenamefont {Bell}\ \emph {et~al.}(2018)\citenamefont {Bell},
  \citenamefont {Rahn},\ and\ \citenamefont {Talbert}}]{Bell:2018vaa}%
  \BibitemOpen
  \bibfield  {author} {\bibinfo {author} {\bibfnamefont {G.}~\bibnamefont
  {Bell}}, \bibinfo {author} {\bibfnamefont {R.}~\bibnamefont {Rahn}}, \ and\
  \bibinfo {author} {\bibfnamefont {J.}~\bibnamefont {Talbert}},\ }\href
  {\doibase 10.1016/j.nuclphysb.2018.09.026} {\bibfield  {journal} {\bibinfo
  {journal} {Nucl. Phys.}\ }\textbf {\bibinfo {volume} {B936}},\ \bibinfo
  {pages} {520} (\bibinfo {year} {2018})},\ \Eprint
  {http://arxiv.org/abs/1805.12414} {arXiv:1805.12414 [hep-ph]} \BibitemShut
  {NoStop}%
\bibitem [{\citenamefont {Bell}\ \emph {et~al.}(2019)\citenamefont {Bell},
  \citenamefont {Rahn},\ and\ \citenamefont {Talbert}}]{Bell:2018oqa}%
  \BibitemOpen
  \bibfield  {author} {\bibinfo {author} {\bibfnamefont {G.}~\bibnamefont
  {Bell}}, \bibinfo {author} {\bibfnamefont {R.}~\bibnamefont {Rahn}}, \ and\
  \bibinfo {author} {\bibfnamefont {J.}~\bibnamefont {Talbert}},\ }\href
  {\doibase 10.1007/JHEP07(2019)101} {\bibfield  {journal} {\bibinfo  {journal}
  {JHEP}\ }\textbf {\bibinfo {volume} {07}},\ \bibinfo {pages} {101} (\bibinfo
  {year} {2019})},\ \Eprint {http://arxiv.org/abs/1812.08690} {arXiv:1812.08690
  [hep-ph]} \BibitemShut {NoStop}%
\bibitem [{\citenamefont {Bell}\ \emph {et~al.}(2020)\citenamefont {Bell},
  \citenamefont {Rahn},\ and\ \citenamefont {Talbert}}]{Bell:2020yzz}%
  \BibitemOpen
  \bibfield  {author} {\bibinfo {author} {\bibfnamefont {G.}~\bibnamefont
  {Bell}}, \bibinfo {author} {\bibfnamefont {R.}~\bibnamefont {Rahn}}, \ and\
  \bibinfo {author} {\bibfnamefont {J.}~\bibnamefont {Talbert}},\ }\href@noop
  {} {\  (\bibinfo {year} {2020})},\ \Eprint {http://arxiv.org/abs/2004.08396}
  {arXiv:2004.08396 [hep-ph]} \BibitemShut {NoStop}%
\bibitem [{\citenamefont {Kardos}\ \emph {et~al.}(2020)\citenamefont {Kardos},
  \citenamefont {Larkoski},\ and\ \citenamefont
  {Tr\'ocs\'anyi}}]{Kardos:2020ppl}%
  \BibitemOpen
  \bibfield  {author} {\bibinfo {author} {\bibfnamefont {A.}~\bibnamefont
  {Kardos}}, \bibinfo {author} {\bibfnamefont {A.~J.}\ \bibnamefont
  {Larkoski}}, \ and\ \bibinfo {author} {\bibfnamefont {Z.}~\bibnamefont
  {Tr\'ocs\'anyi}},\ }\href {\doibase 10.1103/PhysRevD.101.114034} {\bibfield
  {journal} {\bibinfo  {journal} {Phys. Rev. D}\ }\textbf {\bibinfo {volume}
  {101}},\ \bibinfo {pages} {114034} (\bibinfo {year} {2020})},\ \Eprint
  {http://arxiv.org/abs/2002.05730} {arXiv:2002.05730 [hep-ph]} \BibitemShut
  {NoStop}%
\bibitem [{\citenamefont {Hoang}\ \emph {et~al.}(2019)\citenamefont {Hoang},
  \citenamefont {Mantry}, \citenamefont {Pathak},\ and\ \citenamefont
  {Stewart}}]{Hoang:2019ceu}%
  \BibitemOpen
  \bibfield  {author} {\bibinfo {author} {\bibfnamefont {A.~H.}\ \bibnamefont
  {Hoang}}, \bibinfo {author} {\bibfnamefont {S.}~\bibnamefont {Mantry}},
  \bibinfo {author} {\bibfnamefont {A.}~\bibnamefont {Pathak}}, \ and\ \bibinfo
  {author} {\bibfnamefont {I.~W.}\ \bibnamefont {Stewart}},\ }\href {\doibase
  10.1007/JHEP12(2019)002} {\bibfield  {journal} {\bibinfo  {journal} {JHEP}\
  }\textbf {\bibinfo {volume} {12}},\ \bibinfo {pages} {002} (\bibinfo {year}
  {2019})},\ \Eprint {http://arxiv.org/abs/1906.11843} {arXiv:1906.11843
  [hep-ph]} \BibitemShut {NoStop}%
\bibitem [{\citenamefont {Dokshitzer}\ \emph {et~al.}(1997)\citenamefont
  {Dokshitzer}, \citenamefont {Leder}, \citenamefont {Moretti},\ and\
  \citenamefont {Webber}}]{Dokshitzer:1997in}%
  \BibitemOpen
  \bibfield  {author} {\bibinfo {author} {\bibfnamefont {Y.~L.}\ \bibnamefont
  {Dokshitzer}}, \bibinfo {author} {\bibfnamefont {G.~D.}\ \bibnamefont
  {Leder}}, \bibinfo {author} {\bibfnamefont {S.}~\bibnamefont {Moretti}}, \
  and\ \bibinfo {author} {\bibfnamefont {B.~R.}\ \bibnamefont {Webber}},\
  }\href {\doibase 10.1088/1126-6708/1997/08/001} {\bibfield  {journal}
  {\bibinfo  {journal} {JHEP}\ }\textbf {\bibinfo {volume} {08}},\ \bibinfo
  {pages} {001} (\bibinfo {year} {1997})},\ \Eprint
  {http://arxiv.org/abs/hep-ph/9707323} {arXiv:hep-ph/9707323 [hep-ph]}
  \BibitemShut {NoStop}%
\bibitem [{\citenamefont {Wobisch}\ and\ \citenamefont
  {Wengler}(1998)}]{Wobisch:1998wt}%
  \BibitemOpen
  \bibfield  {author} {\bibinfo {author} {\bibfnamefont {M.}~\bibnamefont
  {Wobisch}}\ and\ \bibinfo {author} {\bibfnamefont {T.}~\bibnamefont
  {Wengler}},\ }in\ \href@noop {} {\emph {\bibinfo {booktitle} {{Monte Carlo
  generators for HERA physics. Proceedings, Workshop, Hamburg, Germany,
  1998-1999}}}}\ (\bibinfo {year} {1998})\ pp.\ \bibinfo {pages} {270--279},\
  \Eprint {http://arxiv.org/abs/hep-ph/9907280} {arXiv:hep-ph/9907280 [hep-ph]}
  \BibitemShut {NoStop}%
\bibitem [{\citenamefont {Larkoski}(2020)}]{Larkoski:2020wgx}%
  \BibitemOpen
  \bibfield  {author} {\bibinfo {author} {\bibfnamefont {A.~J.}\ \bibnamefont
  {Larkoski}},\ }\href@noop {} {\  (\bibinfo {year} {2020})},\ \Eprint
  {http://arxiv.org/abs/2006.14680} {arXiv:2006.14680 [hep-ph]} \BibitemShut
  {NoStop}%
\bibitem [{\citenamefont {Catani}\ \emph {et~al.}(1993)\citenamefont {Catani},
  \citenamefont {Trentadue}, \citenamefont {Turnock},\ and\ \citenamefont
  {Webber}}]{Catani:1992ua}%
  \BibitemOpen
  \bibfield  {author} {\bibinfo {author} {\bibfnamefont {S.}~\bibnamefont
  {Catani}}, \bibinfo {author} {\bibfnamefont {L.}~\bibnamefont {Trentadue}},
  \bibinfo {author} {\bibfnamefont {G.}~\bibnamefont {Turnock}}, \ and\
  \bibinfo {author} {\bibfnamefont {B.~R.}\ \bibnamefont {Webber}},\ }\href
  {\doibase 10.1016/0550-3213(93)90271-P} {\bibfield  {journal} {\bibinfo
  {journal} {Nucl. Phys.}\ }\textbf {\bibinfo {volume} {B407}},\ \bibinfo
  {pages} {3} (\bibinfo {year} {1993})}\BibitemShut {NoStop}%
\bibitem [{\citenamefont {Del~Duca}\ \emph
  {et~al.}(2016{\natexlab{a}})\citenamefont {Del~Duca}, \citenamefont {Duhr},
  \citenamefont {Kardos}, \citenamefont {Somogyi},\ and\ \citenamefont
  {Tr\'ocs\'anyi}}]{DelDuca:2016csb}%
  \BibitemOpen
  \bibfield  {author} {\bibinfo {author} {\bibfnamefont {V.}~\bibnamefont
  {Del~Duca}}, \bibinfo {author} {\bibfnamefont {C.}~\bibnamefont {Duhr}},
  \bibinfo {author} {\bibfnamefont {A.}~\bibnamefont {Kardos}}, \bibinfo
  {author} {\bibfnamefont {G.}~\bibnamefont {Somogyi}}, \ and\ \bibinfo
  {author} {\bibfnamefont {Z.}~\bibnamefont {Tr\'ocs\'anyi}},\ }\href {\doibase
  10.1103/PhysRevLett.117.152004} {\bibfield  {journal} {\bibinfo  {journal}
  {Phys. Rev. Lett.}\ }\textbf {\bibinfo {volume} {117}},\ \bibinfo {pages}
  {152004} (\bibinfo {year} {2016}{\natexlab{a}})},\ \Eprint
  {http://arxiv.org/abs/1603.08927} {arXiv:1603.08927 [hep-ph]} \BibitemShut
  {NoStop}%
\bibitem [{\citenamefont {Del~Duca}\ \emph
  {et~al.}(2016{\natexlab{b}})\citenamefont {Del~Duca}, \citenamefont {Duhr},
  \citenamefont {Kardos}, \citenamefont {Somogyi}, \citenamefont {Sz\H{o}r},
  \citenamefont {Tr\'ocs\'anyi},\ and\ \citenamefont
  {Tulip\'ant}}]{DelDuca:2016ily}%
  \BibitemOpen
  \bibfield  {author} {\bibinfo {author} {\bibfnamefont {V.}~\bibnamefont
  {Del~Duca}}, \bibinfo {author} {\bibfnamefont {C.}~\bibnamefont {Duhr}},
  \bibinfo {author} {\bibfnamefont {A.}~\bibnamefont {Kardos}}, \bibinfo
  {author} {\bibfnamefont {G.}~\bibnamefont {Somogyi}}, \bibinfo {author}
  {\bibfnamefont {Z.}~\bibnamefont {Sz\H{o}r}}, \bibinfo {author}
  {\bibfnamefont {Z.}~\bibnamefont {Tr\'ocs\'anyi}}, \ and\ \bibinfo {author}
  {\bibfnamefont {Z.}~\bibnamefont {Tulip\'ant}},\ }\href {\doibase
  10.1103/PhysRevD.94.074019} {\bibfield  {journal} {\bibinfo  {journal} {Phys.
  Rev.}\ }\textbf {\bibinfo {volume} {D94}},\ \bibinfo {pages} {074019}
  (\bibinfo {year} {2016}{\natexlab{b}})},\ \Eprint
  {http://arxiv.org/abs/1606.03453} {arXiv:1606.03453 [hep-ph]} \BibitemShut
  {NoStop}%
\bibitem [{\citenamefont {Tulip\'ant}\ \emph {et~al.}(2017)\citenamefont
  {Tulip\'ant}, \citenamefont {Kardos},\ and\ \citenamefont
  {Somogyi}}]{Tulipant:2017ybb}%
  \BibitemOpen
  \bibfield  {author} {\bibinfo {author} {\bibfnamefont {Z.}~\bibnamefont
  {Tulip\'ant}}, \bibinfo {author} {\bibfnamefont {A.}~\bibnamefont {Kardos}},
  \ and\ \bibinfo {author} {\bibfnamefont {G.}~\bibnamefont {Somogyi}},\ }\href
  {\doibase 10.1140/epjc/s10052-017-5320-9} {\bibfield  {journal} {\bibinfo
  {journal} {Eur. Phys. J.}\ }\textbf {\bibinfo {volume} {C77}},\ \bibinfo
  {pages} {749} (\bibinfo {year} {2017})},\ \Eprint
  {http://arxiv.org/abs/1708.04093} {arXiv:1708.04093 [hep-ph]} \BibitemShut
  {NoStop}%
\bibitem [{\citenamefont {Kardos}\ \emph {et~al.}(2016)\citenamefont {Kardos},
  \citenamefont {Somogyi},\ and\ \citenamefont
  {Tr\'ocs\'anyi}}]{Kardos:2016pic}%
  \BibitemOpen
  \bibfield  {author} {\bibinfo {author} {\bibfnamefont {A.}~\bibnamefont
  {Kardos}}, \bibinfo {author} {\bibfnamefont {G.}~\bibnamefont {Somogyi}}, \
  and\ \bibinfo {author} {\bibfnamefont {Z.}~\bibnamefont {Tr\'ocs\'anyi}},\
  }\bibfield  {booktitle} {\emph {\bibinfo {booktitle} {{Proceedings, 13th DESY
  Workshop on Elementary Particle Physics: Loops and Legs in Quantum Field
  Theory (LL2016): Leipzig, Germany, April 24-29, 2016}}},\ }\href {\doibase
  10.22323/1.260.0021} {\bibfield  {journal} {\bibinfo  {journal} {PoS}\
  }\textbf {\bibinfo {volume} {LL2016}},\ \bibinfo {pages} {021} (\bibinfo
  {year} {2016})}\BibitemShut {NoStop}%
\bibitem [{\citenamefont {Kardos}\ \emph {et~al.}(2018)\citenamefont {Kardos},
  \citenamefont {Somogyi},\ and\ \citenamefont
  {Tr\'ocs\'anyi}}]{Kardos:2018kth}%
  \BibitemOpen
  \bibfield  {author} {\bibinfo {author} {\bibfnamefont {A.}~\bibnamefont
  {Kardos}}, \bibinfo {author} {\bibfnamefont {G.}~\bibnamefont {Somogyi}}, \
  and\ \bibinfo {author} {\bibfnamefont {Z.}~\bibnamefont {Tr\'ocs\'anyi}},\
  }\href {\doibase 10.1016/j.physletb.2018.10.014} {\bibfield  {journal}
  {\bibinfo  {journal} {Phys. Lett.}\ }\textbf {\bibinfo {volume} {B786}},\
  \bibinfo {pages} {313} (\bibinfo {year} {2018})},\ \Eprint
  {http://arxiv.org/abs/1807.11472} {arXiv:1807.11472 [hep-ph]} \BibitemShut
  {NoStop}%
\bibitem [{\citenamefont {Badea}\ \emph {et~al.}(2019)\citenamefont {Badea},
  \citenamefont {Baty}, \citenamefont {Chang}, \citenamefont {Innocenti},
  \citenamefont {Maggi}, \citenamefont {Mcginn}, \citenamefont {Peters},
  \citenamefont {Sheng}, \citenamefont {Thaler},\ and\ \citenamefont
  {Lee}}]{Badea:2019vey}%
  \BibitemOpen
  \bibfield  {author} {\bibinfo {author} {\bibfnamefont {A.}~\bibnamefont
  {Badea}}, \bibinfo {author} {\bibfnamefont {A.}~\bibnamefont {Baty}},
  \bibinfo {author} {\bibfnamefont {P.}~\bibnamefont {Chang}}, \bibinfo
  {author} {\bibfnamefont {G.~M.}\ \bibnamefont {Innocenti}}, \bibinfo {author}
  {\bibfnamefont {M.}~\bibnamefont {Maggi}}, \bibinfo {author} {\bibfnamefont
  {C.}~\bibnamefont {Mcginn}}, \bibinfo {author} {\bibfnamefont
  {M.}~\bibnamefont {Peters}}, \bibinfo {author} {\bibfnamefont {T.-A.}\
  \bibnamefont {Sheng}}, \bibinfo {author} {\bibfnamefont {J.}~\bibnamefont
  {Thaler}}, \ and\ \bibinfo {author} {\bibfnamefont {Y.-J.}\ \bibnamefont
  {Lee}},\ }\href {\doibase 10.1103/PhysRevLett.123.212002} {\bibfield
  {journal} {\bibinfo  {journal} {Phys. Rev. Lett.}\ }\textbf {\bibinfo
  {volume} {123}},\ \bibinfo {pages} {212002} (\bibinfo {year} {2019})},\
  \Eprint {http://arxiv.org/abs/1906.00489} {arXiv:1906.00489 [hep-ex]}
  \BibitemShut {NoStop}%
\bibitem [{\citenamefont {Aaboud}\ \emph {et~al.}(2018)\citenamefont {Aaboud}
  \emph {et~al.}}]{Aaboud:2017qwh}%
  \BibitemOpen
  \bibfield  {author} {\bibinfo {author} {\bibfnamefont {M.}~\bibnamefont
  {Aaboud}} \emph {et~al.} (\bibinfo {collaboration} {ATLAS}),\ }\href
  {\doibase 10.1103/PhysRevLett.121.092001} {\bibfield  {journal} {\bibinfo
  {journal} {Phys. Rev. Lett.}\ }\textbf {\bibinfo {volume} {121}},\ \bibinfo
  {pages} {092001} (\bibinfo {year} {2018})},\ \Eprint
  {http://arxiv.org/abs/1711.08341} {arXiv:1711.08341 [hep-ex]} \BibitemShut
  {NoStop}%
\bibitem [{\citenamefont {Sirunyan}\ \emph {et~al.}(2018)\citenamefont
  {Sirunyan} \emph {et~al.}}]{Sirunyan:2018xdh}%
  \BibitemOpen
  \bibfield  {author} {\bibinfo {author} {\bibfnamefont {A.~M.}\ \bibnamefont
  {Sirunyan}} \emph {et~al.} (\bibinfo {collaboration} {CMS}),\ }\href
  {\doibase 10.1007/JHEP11(2018)113} {\bibfield  {journal} {\bibinfo  {journal}
  {JHEP}\ }\textbf {\bibinfo {volume} {11}},\ \bibinfo {pages} {113} (\bibinfo
  {year} {2018})},\ \Eprint {http://arxiv.org/abs/1807.05974} {arXiv:1807.05974
  [hep-ex]} \BibitemShut {NoStop}%
\bibitem [{\citenamefont {Aad}\ \emph {et~al.}(2019)\citenamefont {Aad} \emph
  {et~al.}}]{Aad:2019vyi}%
  \BibitemOpen
  \bibfield  {author} {\bibinfo {author} {\bibfnamefont {G.}~\bibnamefont
  {Aad}} \emph {et~al.} (\bibinfo {collaboration} {ATLAS}),\ }\href@noop {} {\
  (\bibinfo {year} {2019})},\ \Eprint {http://arxiv.org/abs/1912.09837}
  {arXiv:1912.09837 [hep-ex]} \BibitemShut {NoStop}%
\end{thebibliography}%

\end{document}